\documentclass[ aps,twocolumn, floatfix, showpacs]{revtex4}
\usepackage{graphicx,epsfig,amsmath,amsfonts,epic,eepic,psfrag,latexsym,
amssymb}

\begin{document}

\title{Kinetics of cell division in epidermal maintenance}
\author{Allon M. Klein$^1$, David P. Doup\'e$^2$, Phillip H. Jones$^2$, and Benjamin D. Simons$^1$}
\affiliation{$^1$Cavendish Laboratory, Madingley Road, Cambridge CB3 OHE, UK\\
$^2$ MRC Cancer Cell Unit, Hutchison-MRC Research Centre, Cambridge CB2 2XZ, 
UK}


\begin{abstract}
The rules governing cell division and differentiation are central to 
understanding the mechanisms of development, aging and cancer. 
By utilising inducible genetic labelling, recent studies have shown 
that the clonal population in transgenic mouse epidermis can be 
tracked \emph{in vivo}. Drawing on these results, we explain how 
clonal fate data may be used to infer the rules
of cell division and 
differentiation underlying the maintenance of adult murine tail-skin. 
We show that the rates of cell division and differentiation may be 
evaluated by considering the long-time and short-time clone fate 
data, and that the data is consistent with cells dividing 
independently rather than synchronously. Motivated by these findings, 
we consider a mechanism for cancer onset based closely on the model 
for normal adult skin. By analysing the expected changes to clonal
fate in cancer emerging from a  simple two-stage mutation, we propose 
that clonal fate data may provide a novel method for studying the 
earliest stages of the disease.
\end{abstract}

\maketitle

\section{Introduction}
\begin{figure}[t]
\begin{center}
\includegraphics[width=3.3in]{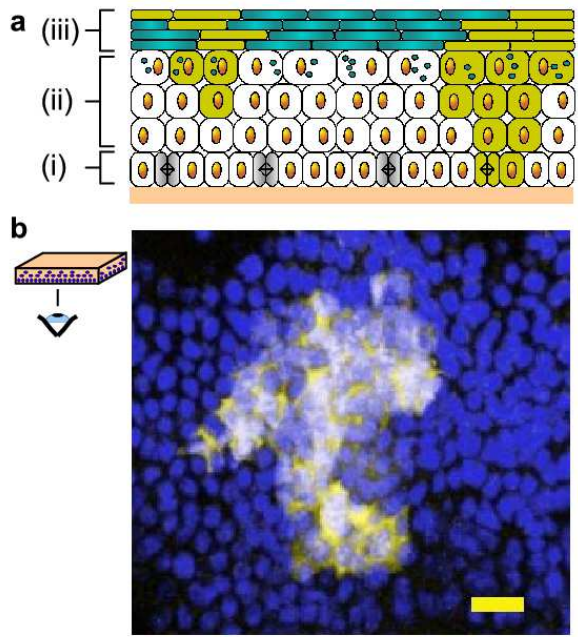}
\caption{\small 
(Color online) 
(a) Schematic cross-section of murine interfollicular epidermis (IFE)
 showing the organisation of cells within different layers and 
indicating the 
architecture of typical labelled clones. Proliferating cells (grey) 
are confined to the basal layer (labelled i); differentiated cells 
migrate through the superbasal layers (ii), where they flatten 
into cornified cells, losing their nuclei and assembling a 
cornified envelope (green) (iii), eventually becoming shed 
at the surface. 
The shaded regions (yellow) indicate two distinct clones, the 
progeny of single basal layer cells labelled at induction. While the clone on 
the right retains at least one labelled cell in the basal layer, the clone on
the left hand side has detached from the basal layer indicating that all of 
the cells have stopped proliferating. 
The former are designated as ``persisting clones'' and contribute to the clone 
size distributions, while the latter, being difficult to resolve reliably, are 
excluded from experimental consideration. 
(b) Typical example of a clone acquired at a late time point, viewed from the basal layer surface. Cell nuclei are labelled blue; the 
hereditary clone marker (EYFP) appears yellow. Scale bar: $20\mu$m}
\label{fig:skinCrossSection}
\end{center}
\end{figure}
A major challenge in biology is to determine how proliferating cells in 
developing and adult tissues behave \emph{in vivo}. A powerful technique in 
solving this problem is clonal analysis, the labelling of a sample of cells 
within the tissue to enable their fate and that of their progeny to be 
tracked~\cite{Clarke99}. This approach gives access to information on 
proliferation, migration, differentiation (into other cell types), and cell 
death (apoptosis) of the labelled cell population. The most reliable method 
of labelling is through genetic modification leading to the expression
of a reporter gene in a random sample of cells. Recently it has become 
possible to activate genetic labelling at a defined time in transgenic 
mice, enabling the kinetics of labelled cells to be studied with single-cell 
resolution \emph{in vivo}~\cite{Jonkers02}. From a theoretical perspective, 
the analysis of clonal fate data presents a challenging ``inverse problem'' in 
population dynamics: While it is straightforward to predict 
the time-evolution of a population distribution 
according to a set of growth rules, the analysis of the inverse problem is 
more challenging, open to ambiguity and potential misinterpretation. 

These principles are exemplified by the mechanism of murine epidermal 
homeostasis: Mammalian epidermis is organised into hair follicles 
interspersed with \emph{interfollicular}
epidermis (IFE), 
which consists of layers of specialised cells known as keratinocytes 
\cite{Fuchs:07} (see Fig.~\ref{fig:skinCrossSection}(a)). Proliferating cells are confined to the basal epidermal layer. As they differentiate into specialised skin cells, the basal cells withdraw from the cycle of cell proliferation and then leave the basal layer, migrating towards the epidermal surface from which they are ultimately shed. To
maintain the integrity of the tissue, new cells must be generated to replace those lost through shedding. For many years, it has been thought that interfollicular epidermis is maintained by two distinct progenitor cell populations in the basal layer. These comprise long-lived stem cells (S) with the capacity to self-renew, and their progeny, known as transit-amplifying cells (TA), which
go on to differentiate and exit the basal layer after several rounds of cell division \cite{Potten:81}.  Stem cells are also found in the hair follicles, but whilst they have the potential to generate epidermis in circumstances such as wounding, they do not appear to contribute to maintaining normal epidermis \cite{Ito:05, Levy:05}.

The prevailing model of interfollicular homeostasis posits that 
the tissue is organised into regularly sized ``epidermal proliferative 
units'' or EPUs, in which a central stem cell supports a surrounding, clonal,
 population of transit amplifying cells, which in turn generate a column of overlying differentiated cells \cite{Mackenzie:70, Potten:74}. Several experimental approaches have been used to attempt to demonstrate the existence of EPUs, but conclusive evidence for their existence is lacking.  The EPU model predicts that slowly-cycling stem cells should be found in a patterned array in the IFE; cell labelling studies have failed to demonstrate such a pattern \cite{Braun03}.  In chimaeric mice the EPU model predicts that the boundaries of mosaicism in the IFE should run along the boundaries of EPUs; instead boundaries were found to be highly irregular \cite{Schmidt:87}.  Genetic labelling studies using viral infection or mutation to activate expression of a reporter gene in epidermal cells have demonstrated the existence of long-lived, cohesive clusters of labelled cells in the epidermis, but these clusters do not conform to the predicted size distribution of the EPU \cite{Braun03, Ghazizadeh:01, Kameda:03, Potten:81, Ro:04, Ro:05}.  

Thus, until recently the means by which homeostasis of IFE was achieved has been unclear.  
However, by exploiting inducible genetic labelling, recent studies have 
allowed the fate of a representative sample of progenitor cells and their 
progeny to tracked \emph{in vivo}~\cite{Clayton:07}. As well as undermining the basis of the 
stem/TA cell hypothesis, the range of clone fate data provide the means to 
infer the true mechanism of epidermal homeostasis. In 
particular, these investigations indicate that the maintenance of IFE in the 
adult system conforms to a remarkably simple birth-death process involving a 
single progenitor cell compartment. Expanding upon the preliminary theoretical 
findings of Ref.~\cite{Clayton:07}, the aim of this paper is to elucidate in 
full the evidence for, and the properties of, the model of epidermal 
maintenance, and to describe the potential of the system as a method to 
explore early signatures of carcinogenic mutations.

\subsection{Background: Experimental Methodology}
To organise our discussion, we begin with an overview of the 
experimental arrangement, referring to Ref.~\cite{Clayton:07} for technical details of the experimental system. To generate data on the fate of individual labelled cells and their progeny, hereafter referred to as clonal fate data, inducible genetic marking was used to label a sample of cells and their progeny in the epidermis of transgenic mice.  The enhanced Yellow Fluorescent Protein (EYFP) label was then detected by confocal microscopy, which enables 3D imaging of entire sheets of epidermis. Low-frequency labelling of approximately 1 in 600 basal-layer epidermal cells at a defined time was achieved by using two drugs to mediate a genetic event which resulted in expression of the EYFP gene in a cohort of mice.  This low efficiency labelling ensures that clones are unlikely to merge (see discussion in section~\ref{sec:model}).  
By analysing samples of mice at different time points it was possible to 
analyse the fate of labelled clones at single cell resolution \emph{in vivo} 
for times up to one year post-labelling in the epidermis 
(see, for example, Fig.~\ref{fig:skinCrossSection}(b))
\cite{Braun03, Clayton:07}.
\begin{figure}[t]
\begin{center}
\includegraphics[width=3.3in]{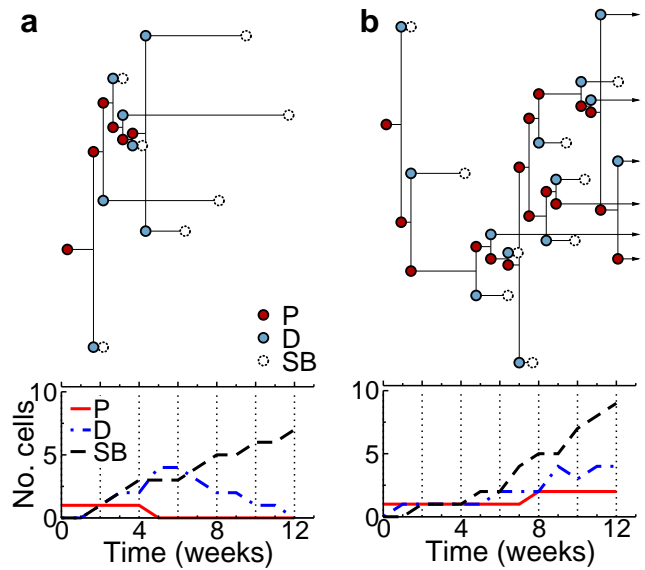}
\caption{\small 
(Color online)
\emph{Top:} Theoretical lineage for the first $12$ weeks post-labelling 
of (a) a detached clone in which all
cells have undergone a transition to terminal differentiation by week $12$, and (b) a 
persisting clone in which some of the cells maintain a proliferative 
capacity, according to model (\ref{modelRateLaws}). Circles indicate progenitor cells (P), differentiated cells (D), and suprabasal cells (SB). Note that, because the birth-death process (\ref{modelRateLaws}) is Markovian, the lifetime of cells is drawn from a Poisson distribution with no strict minimum or maximum lifetime. The statistics of such lineage trees do not change significantly when we account for a latency period between divisions that is much shorter than the mean cell lifetime (see discussion in section \ref{sec:stochastic}).
\emph{Bottom:} The total number of proliferating, differentiated and supra-basal cells for the two clones as a function of time.}
\label{fig:tree}
\end{center}
\end{figure}

With the gradual accummulation of EYFP levels, the early time data (less than
two weeks) reveals a small increase in the number of labelled clones containing
one or two cells. At longer times, clones increase in size while cells within
clones begin to migrate through the suprabasal layers forming relatively 
cohesive irregular columns (see Fig.~\ref{fig:skinCrossSection}(a)).  

The loss of nuclei in the cornified layer (fig.~\ref{fig:skinCrossSection}) makes determination of the number of 
cornified layer cells in larger clones by microscopy unreliable. 
Therefore, to identify a manageable population, attention was focused on 
the population of basal cells in
``persisting clones'', defined as those labelled clones which retain 
at least one basal layer cell, such as is exemplified in the theoretical 
lineage maps in Fig.~\ref{fig:tree}. 
After two weeks, the density of persisting clones was seen to decrease 
monotonically indicating that the entire cell population within such clones 
had become differentiated and the clone detached from the basal layer 
(shown schematically in figs.~\ref{fig:skinCrossSection}(a) and \ref{fig:tree}(a)). 
However, the population of persisting clones showed a steady increase in 
size throughout the entire duration of the experiment. 
\begin{figure}
\begin{center}
\includegraphics[width=3.3in]{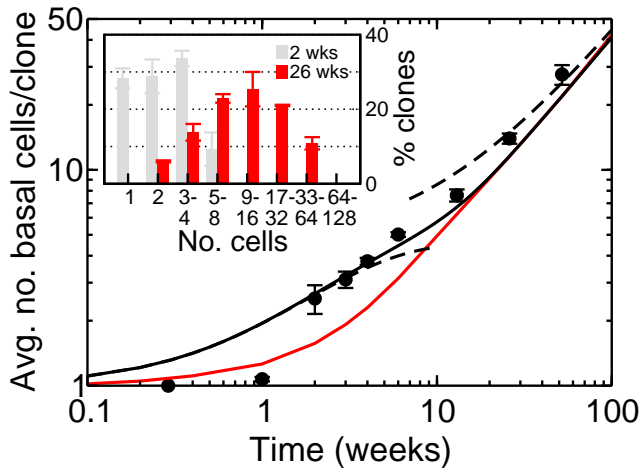}
\caption{\small
(Color online)
Mean number of basal layer cells in persisting clones. The experimental data (circles) 
show an inexorable increase in the size of persisting clones over the entire
time course of the experiment. The behaviour at short times (from $2-6$ weeks) and at long times (beyond $13$ weeks) follows the two simple analytical approximations described in the main text (lower and upper dashed curves).
For times earlier than two weeks (referring to section~\ref{sec:EYFP}), clones remain approximately 
one cell in size. 
The experimental data are consistent with the behaviour 
predicted by process (\ref{modelRateLaws})
(black line) when it is assumed that only A-type cells 
are labelled at induction. In contrast, assuming that A and B type cells label in 
proportion to their steady-state population leads
to an underestimate of average clone size between two and six weeks (lower curve, red 
online), as does the assumption that type B cells label with better efficiency 
(not shown).\ \ 
\emph{Inset:} The underlying distribution of basal cells per clone at 2 weeks and 26 weeks post-labelling. The data is binned by cell count in increasing powers of $2$.}
\label{fig:avgSize}
\end{center}
\end{figure}

To what extent are the clone fate data consistent with the orthodox stem/TA 
cell model of epidermal maintenance? Referring to Fig.~\ref{fig:avgSize}, 
one observes an inexorable increase in the average size of an 
ever-diminishing persisting clone population. This result is incompatible 
with any model in which the IFE is supported by a population of long-lived 
stem cells. With the latter, one would expect the number density of 
persisting clones to reach a non-zero minimum (commensurate with the labelling 
frequency of stem cells) while the average clone size would asymptote to a 
constant value characteristic of a single epidermal proliferative unit. We 
are therefore lead to abandon, or at least substantially revise, the orthodox 
stem/TA cell hypothesis and look for a different paradigm for epidermal maintenance.

But, to what extent are the clone fate data amenable to theoretical analysis?
Indeed, the application of population dynamics to the problem of cell kinetics 
has a long history (see, e.g., Refs.\cite{Loeffler:80, Loeffler:91,White:00,%
Roeder:06}) with studies of epidermal cell proliferation addressed in 
several papers~\cite{Savill:03,Appleton:77,Potten:82,Weinstein:84}. However,
even in the adult system, where cell kinetics may be expected to conform to a 
``steady-state'' behaviour, it is far from clear whether the cell dynamics can 
be modelled as a simple stochastic process. Regulation due to environmental 
conditions could lead to a highly nonlinear or even non-local dependence of 
cell division rates. Indeed, \emph{a priori}, it is far from clear whether 
the cell kinetics can be considered as Markovian, i.e. that cell division 
is both random and independent of the past history of the cell. Therefore,
instead of trying to formulate a complex theory of cell division, taking 
account of the potential underlying biochemical pathways and regulation 
networks~\cite{Savill:03}, we will follow a different strategy looking for 
signatures of steady-state behaviour in the experimental data and evidence 
for a simple underlying mechanism for cell fate. Intriguingly, such evidence 
is to be found in the scaling properties of the clone size 
distribution~\cite{Clayton:07}.

\subsection{Scaling}
\label{sec:scaling}
\begin{figure}
\begin{center}
\includegraphics[width=3.3in]{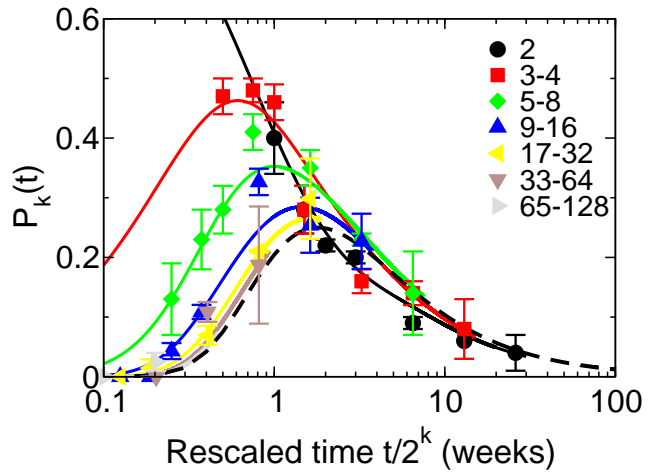}
\caption{\small 
(Color online)
Time dependence of the grouped size distribution of 
\emph{persisting} clones, $\mathcal{P}^{\rm pers.}_{k}(t)$, plotted as a 
function of the rescaled time coordinate $t/2^k\mapsto t$. 
The data points show measurements 
(extracted from data such as shown in Fig.~\ref{fig:avgSize}(inset), 
given fully in ref.~\cite{Clayton:07}),
while the solid curves show the probability 
distributions associated with the non-equilibrium 
process~(\ref{modelRateLaws}) for the basal-layer clone population as 
obtained by a numerical solution of the Master equation~(\ref{pMasterEqn}).
(Error bars refer to standard error of the mean). 
At long times, the data 
converge onto a universal curve (dashed line), which one may identify with 
the form given in eq.~\ref{PK_Exact_All_Cells}. The rescaling compresses the 
time axis for larger clones, so that the large-clone distributions appear to 
converge much earlier onto the universal curve.}
\label{fig:scaling}
\end{center}
\end{figure}

To identify scaling characteristics, it is necessary to focus on the basal 
layer clone size distribution, $P_n(t)$, which describes the probability that 
a labelled progenitor cell develops into a clone with a total of $n$ basal 
layer cells at a time $t$ after labelling. (Note that, in general, the total 
number of cells in the supra-basal layers of a clone may greatly exceed the 
number of basal layer cells.) With this definition, $P_0(t)$ describes the 
``extinction'' probability of a clone, i.e. the probability that \emph{all} 
of the cells within a labelled clone have migrated into the supra-basal 
layers. To make contact with the experimental data, it is necessary to 
eliminate from the statistical ensemble the extinct clone population (which 
are difficult to monitor experimentally) and single-cell clones (whose 
contribution to the total ensemble is compromised by the seemingly unknown 
relative labelling efficiency of proliferating and post-mitotic cells at 
induction), leading to a reduced distribution for ``persisting" clones, 
\begin{eqnarray*}
P^{\rm pers.}_{n\ge 2}(t)\equiv\frac{P_n(t)}{1-P_0(t)-P_1(t)}\,.
\end{eqnarray*}
Then, to consolidate the data and minimise fluctuations due to counting 
statistics, it is further convenient to \emph{bin} the distribution in 
increasing powers of 2,
\begin{eqnarray*} 
\mathcal{P}^{\rm pers.}_{k}(t) = \sum_{n=2^{k-1}+1}^{2^k} 
P^{\rm pers.}_{n\geq2}(t)\,,
\end{eqnarray*}
i.e. $\mathcal{P}^{\rm pers.}_{1}(t)$ describes the probability of having 
two cells per clone, $\mathcal{P}^{\rm pers.}_{2}(t)$ describes the probability
of having 3-4 cells per clone, and so on. Referring to 
Fig.~\ref{fig:scaling}, one may see that, after an initial transient 
behaviour, the clone size distribution asymptotes in time to the simple 
scaling form,
\begin{eqnarray}
\mathcal{P}^{\rm pers.}_{k}(t)= f(2^k/t)\,.
\label{scaling}
\end{eqnarray}

This striking observation brings with it a number of important consequences:
As well as reinforcing the inapplicability of the stem cell/TA cell 
hypothesis, such behaviour suggests that epidermal maintenance must conform 
to a simple model of cell division. The absence of further characteristic 
time-scales, beyond that of an overall proliferation rate, motivates the 
consideration of a simple kinetics in which \emph{only one process dictates 
the long-time characteristics of clonal evolution}. 

Moreover, from the scaling observation one can also deduce two additional 
constraints: Firstly, in the long-time limit, the average number of basal
layer cells within a \emph{persisting} clone \emph{increases linearly with 
time}, viz. 
\begin{eqnarray*}
P^{\rm pers.}_{n\ge 2}(t) &\simeq& \frac{d}{d(2^k)} 
\mathcal{P}^{\rm pers.}_{k}(t) =  \frac{1}{t}\, f' (2^k / t) \\
\langle n\rangle_{\rm pers.} &\equiv & \sum_{n\ge 2}^\infty n\, 
P^{\rm pers.}_{n\ge 2}(t)\simeq \int_0^\infty dn \frac{n}{t} \, 
f'(n/t)
\propto t\,.
\end{eqnarray*}
Secondly, if we assume that labelled progenitor cells are representative of 
\emph{all} progenitor cells in the epidermis, and that the population of 
clones with only one basal layer cell is not ``extensive'' (i.e. 
$\lim_{t\to\infty} P_1(t)=0$), this means that, in the long-time limit, the 
clone persistence probability must scale as $1-P_0(t)\propto 1/t$ such that 
\begin{eqnarray*}
\langle n\rangle=\sum_n n\, P_n(t)\stackrel{!}{=} \rho\,,
\end{eqnarray*}
where the constant, $\rho$, is given by the fraction of proliferating cells 
in the basal layer. Without this condition, one is lead to conclude that the 
labelled population of basal layer cells either grows or diminishes, a 
behaviour incompatible with the (observed) steady-state character of the 
adult system.

Although the manifestation of scaling behaviour in the clone size distributions
gives some confidence that the mechanism of cell fate in IFE conforms to a 
simple non-equilibrium process, it is nevertheless possible to conceive of 
complicated, multi-component, models which could asymptote to the same 
long-time evolution. To further constrain the possible theories, it is helpful 
to draw on additional experimental observations~\cite{Clayton:07}: Firstly, 
immunostaining of clones with a total of two cells (using the proliferation 
marker Ki67 and, separately, the replication licensing factor cdc6) reveals 
that a single cell division may generate either one proliferating and one 
non-proliferating daughter through asymmetric division, or two proliferating 
daughters, or two non-proliferating daughters (cf.~\cite{Lechler,Smart,%
Zhong}). 
Secondly, three-dimensional imaging of the epidermis reveals that only $3\%$ 
of mitotic spindles lie perpendicular to the basal layer indicating that 
divisions may be considered to be confined to the basal layer, 
confirming the results of earlier work that indicates a dividing basal cell 
generates two basal layer cells~\cite{Smart}.

This completes our preliminary discussion of the experimental background and 
phenomenology. In summary, the clone fate data reveal a behaviour wholely 
incompatible with any model based on the concept of long-lived self-renewing 
stem cells. The observation of long-time scaling behaviour motivates the 
consideration of a simple model based on a stochastic non-equilibrium process 
and is indicative of the labelled cells being both a representative (i.e. 
self-sustaining) population and in steady-state. In the following, we will 
develop a theory of epidermal maintenance which encompasses all of these 
observations.

\section{Theory of epidermal maintenance}

\subsection{Model}
\label{sec:model}
Taken together, the range of clonal fate data and the observation of symmetric
and asymmetric division are consistent with a remarkably simple model of 
epidermal homeostasis involving only one proliferating cell compartment and 
engaging just three adjustable parameters: the overall cell division rate, 
$\lambda$; the proportion of cell divisions that are symmetric, $r$; and the 
rate of transfer, $\Gamma$, of non-proliferating cells from the basal to the 
supra-basal layers. To maintain the total proliferating cell population, a 
constraint imposed by the steady-state assumption, we have used the fact 
that the division rates associated with the two channels of symmetric cell 
division must be equal. Denoting the proliferating cells as type A, 
differentiated basal layer cells as type B, and supra-basal layer cells as 
type C, the model describes the non-equilibrium process,
\begin{equation}
\label{modelRateLaws}
\begin{array}{lcl}
&&{\rm A}\stackrel{\lambda}{\longrightarrow}
        \left\{\begin{array}{cl}
        {\rm A}+{\rm A} & {\rm Prob.\ }r \\
        {\rm A}+{\rm B} &  {\rm Prob.\ }1-2r\\
        {\rm B}+{\rm B} &  {\rm Prob.\ }r 
        \end{array}\right. \\
&&{\rm B}\stackrel{\Gamma}{\longrightarrow}{\rm C}\,.\\
\end{array}
\end{equation}
Finally, the experimental observation that the total basal layer cell 
density remains approximately constant over the time course of the experiment 
leads to the additional constraint that 
\begin{eqnarray*}
\Gamma=\frac{\rho}{1-\rho}\lambda\,,
\end{eqnarray*}
reducing the number of adjustable parameters to just two. 

By ignoring processes involving the shedding of cells from the surface of the 
epidermis, the applicability of the model to the consideration
of the \emph{total} clone size distribution is limited to appropriately short 
time scales (up to six weeks post-labelling). However, if we focus only on 
the clone size distribution associated with those cells which occupy the basal 
layer, the model can be applied up to arbitrary times. In this case, the 
transfer process must be replaced by one in which 
${\rm B}\stackrel{\Gamma}{\longrightarrow}\emptyset$. In either case, 
if we treat all instances of cell division and cell transfer as independent stochastic events, a point that we shall revisit later, then the 
time evolution associated with the process~(\ref{modelRateLaws}) can be 
cast in the form of a Master equation. Defining $P_{n_{\rm A},n_{\rm B}}(t)$ as the 
probability of finding $n_{\rm A}$ type A cells and $n_{\rm B}$ type B cells in a given 
clone after some time $t$, the probability distribution evolves according to 
the Master equation:
\begin{eqnarray}
\label{pMasterEqn}
&&\partial_t P_{n_{\rm A},n_{\rm B}} = r\lambda\left[(n_{\rm A}-1)P_{n_{\rm A}-1,n_{\rm B}}-n_{\rm A} P_{n_{\rm A},n_{\rm B}}
\right]\nonumber\\
&&\qquad\qquad + r\lambda[(n_{\rm A}+1)P_{n_{\rm A}+1,n_{\rm B}-2}-n_{\rm A} P_{n_{\rm A},n_{\rm B}}]\nonumber\\
&&\qquad\qquad + (1-2r)\lambda[n_{\rm A} P_{n_{\rm A},n_{\rm B}-1}-n_{\rm A} P_{n_{\rm A},n_{\rm B}}]\nonumber\\
&&\qquad\qquad + \Gamma[(n_{\rm B}+1)P_{n_{\rm A},n_{\rm B}+1}-n_{\rm B} P_{n_{\rm A},n_{\rm B}}]\,.
\end{eqnarray}
If we suppose that the basal layer cells label in proportion to their 
population, the latter must be solved subject to the boundary condition 
$P_{n_{\rm A},n_{\rm B}}(0)=\rho\delta_{n_{\rm A},1}\delta_{n_{\rm B},0}+(1-\rho)\delta_{n_{\rm A},0}
\delta_{n_{\rm B},1}$. Later, in section~\ref{sec:EYFP}, we will argue that the 
clone size distribution is compatible with a labelling efficiency which 
favours A over B type cells. Either way, by excluding single cell clones
from the distribution, this source of ambiguity may be safely eliminated. 
Although the Master equation (and its total cell number generalisation) is 
not amenable to exact analytic solution, its properties can be inferred from 
the consideration of the A cell population alone for which an explicit solution
may be derived. 

When considered alone, A type cells conform to a simple set of rate laws,
\begin{equation}
\label{modelRateLaws2}
\begin{array}{lcl}
&&{\rm A}\stackrel{2r\lambda}{\longrightarrow}
        \left\{\begin{array}{cl}
        {\rm A}+{\rm A} & {\rm Prob.\ }1/2\,, \\
        \emptyset &  {\rm Prob.\ }1/2\,,\\
        \end{array}\right. 
\end{array}
\end{equation}
an example of a Galton-Watson process, long known to statisticians (see, e.g., 
Ref.~\cite{Bailey}). In this case, the probability distribution, which is 
related to that of the two-component model through the relation 
$p_{n_{\rm A}}(t)=\sum_{n_{\rm B}=0}^\infty P_{n_{\rm A},n_{\rm B}}(t)$, 
can be solved analytically. (Here, we have used a lower case $p$ to 
discriminate the probability distribution from its two-component counterpart.)
For an initial distribution $p_{n_{\rm A}}(0)=\delta_{n_{\rm A},1}$ it may be
shown that~\cite{Bailey}, 
\begin{equation}
\label{P_Exact_ACells}
p_{n_{\rm A}}(t) =\left(1+\frac{1}{r\lambda t}\right)^{-(n_{\rm A}+1)} 
\times \left\{ \begin{array}{ll} 
1& n_{\rm A} =0\,,\\ 
\frac{1}{(r\lambda t)^2} & n_{\rm A} > 0\,.
\end{array}\right. 
\end{equation}

From this system and its associated dynamics, one can draw several key 
implications:

\subsubsection{Epidermis is maintained through an ever-decreasing clonal 
population:}

Starting with a single labelled cell, the Galton-Watson process 
predicts that the persistance probability of the resulting clone (i.e., in 
this case, the probability that the clone retains at least one proliferating 
cell), is given by 
\begin{eqnarray*}
p_{n_{\rm A}>0}\equiv 1-p_{0}(t)=\frac{1}{1+r\lambda t}\,,
\end{eqnarray*}
i.e. as with the experiment, the persistance probability of a clone decays 
monotonically, asymptoting to the form $1-p_{0}(t)\propto 1/t$ at time scales 
$t\gg 1/r\lambda$, the time scale for symmetric division. Applied to the 
experimental system, this suggests that labelled clones continue to detach 
from the basal layer indefinitely. At the same time, defining 
\begin{eqnarray*}
p^{\rm pers.}_{n_{\rm A}>0}(t)=\frac{p_{n_{\rm A}}(t)}{1-p_0(t)}\,,
\end{eqnarray*}
as the size distribution of \emph{persisting} clones, the mean number of 
basal layer cells in a persisting clone grows steadily as
\begin{eqnarray*}
\langle n_{\rm A}\rangle_{\rm pers.} \equiv \sum_{n=1}^\infty n_{\rm A}\, 
p^{\rm pers.}_{n_{\rm A}>0}(t) = 1 + r\lambda t\,,
\end{eqnarray*}
such that the overall cell population remains constant, viz. $\langle n_{\rm A}
\rangle\equiv \sum_{n=0}^\infty n_{\rm A}\, p_{n_{\rm A}}(t)=1$, i.e. the continual 
extinction of clones is compensated by the steady growth of persisting clones 
such that the average number of proliferating cells remains constant: given 
enough time, all cells would derive from the same common ancestor, the 
hallmark of the Galton-Watson process~\footnote{Curiously, Galton and Watson 
first dealt with this question amidst the concern that aristocratic surnames 
were becoming ``extinct'' in Victorian Britain~\cite{GaltonWatson}!}.

This linear increase in clone size may lead one to worry about neighbouring 
clones coalescing. Fortunately, the continual extinction of clones ensures 
that the fraction of clones conjoined with their neighbours remains small and 
of same order as the initial labelling density~\footnote{The fraction of 
  clones in contact with their neighbours is estimated by assuming that all 
  clones are randomly and independently distributed, and by noting the 
  empirical observation that clones are oblique in shape, so that the 
  distribution in clone area has the same form as $P^{\rm pers.}_{n>0}(t)$.}. 
The fact that this fraction is constant is again indicative of the 
steady-state condition maintained throughout the experiment.

\subsubsection{Larger clones begin to exhibit the stability of the macroscopic 
system:}

If, at some instant, a clone is seen to have, say, $N_A$ 
proliferating cells then, after a further time $t$, its size will fluctuate 
as
\begin{eqnarray*}
\frac{\langle (n_{\rm A}-\langle n_{\rm A} \rangle)^2\rangle^{1/2}}{\langle n_{\rm A} \rangle}
=\sqrt\frac{2r\lambda t}{N_A}\,.
\end{eqnarray*}
Thus clones (as defined by the A cell population) will maintain an 
approximately stable number of cells providing $t\ll N_A/r\lambda$. For larger 
clones this time may exceed the lifetime of the system. At the limit where 
macroscopic sections of the basal layer are considered, the statistical 
fluctuations are small. 
The increased stability of larger clones also explains the surprising 
prediction that, given enough time, all clones eventually become extinct 
(viz. $\lim_{t\to\infty} p_{n>0}(t)=0$). Calculated explicity, the 
extinction probability for a clone of size $N_A\gg 1$ scales as $p_{0}(t) 
\approx e^{-N_A/r\lambda t}$~\cite{Bailey} approaching unity at long times. 
However, because this extinction probability is small when $ t \ll N_A/
r\lambda$, a large enough clone may easily persist beyond the lifetime of 
the system.



\subsubsection{The properties of the proliferating cell population dictates 
the behaviour of the entire clone size distribution:}


At asymptotically long times, one may show~\footnote{To obtain 
  Eq.~(\ref{P_Exact_All_Cells}), we 
  treat $n_{\rm A},\ n_{\rm B}$ as continuous variables in Eq.~(\ref{pMasterEqn}) (a good 
  approximation at large values of $n$). Then, making the ansatz that the 
  B-cell population remains slave to the A-cell population viz. $n_{\rm A} 
  = \rho n,\ n_{\rm B} = (1-\rho)n$, the Master equation simplifies to 
  the approximate form $$\partial_t P = \frac{r\lambda}{\rho}(n\partial_n^2 P 
  + 2\partial_n P),$$ which is solved by $P_n(t) = ({\rho}/{r\lambda t})^2
  \exp(-{\rho}n/{r\lambda t})$, leading to Eq.~\ref{P_Exact_All_Cells}.}
that the full probability distribution for finding $n=n_{\rm A}+n_{\rm B}$ cells within 
a persisting clone scales in proportion to $p^{\rm pers.}_{n_{\rm A}}(t)$, viz.
\begin{equation}
\label{P_Exact_All_Cells}
\lim_{t \gg 1/r\lambda}P^{\rm pers.}_{n>0}(t)=\frac{\rho}{r\lambda t} 
\exp\left[-\frac{\rho n}{r\lambda t}\right]\, ,
\end{equation}
and so
\begin{equation}
\label{PK_Exact_All_Cells}
\lim_{t \gg 1/r\lambda}\mathcal{P}^{\rm pers.}_{k}(t) \simeq
\exp\left[-2^{k}\frac{\rho}{2r\lambda t}\right] - \exp\left[-2^k\frac{\rho}{r\lambda t}\right]\, ,
\end{equation}
i.e. the probability distribution acquires the scaling form found empirically.
Referring to Eq.~(\ref{scaling}), we can therefore deduce the form of the 
scaling function, 
\begin{equation}
f(x)=\exp[-\rho x/2r\lambda]-\exp[-\rho x/r\lambda]\,. 
\end{equation}
As a result, at long times,
the average basal layer population of persisting clones becomes proportional 
to the average number of proliferating cells per clone, $\langle n\rangle_{\rm 
pers.}  = (1 + r\lambda t)/\rho$, a behaviour consistent with that seen in 
experiment (see Fig.~\ref{fig:avgSize}).



\subsubsection{The creation and transfer of differentiated cells dictates the 
short-time behaviour of the clone size distribution:}


In fitting the model to the data (see below), we will find that the rates 
$\lambda$ and $\Gamma$ at which differentiated cells are created and then 
transferred into the super-basal region are significantly larger than the 
rate of symmetric division $r\lambda$, which dictates the long-time behaviour 
of the clone size distribution. In this case, at early times ($t \lesssim 
1/\Gamma$), the clone size distributions are dominated by the differentiation 
and transfer rates, which remain prominent until the population of labelled 
differentiated cells associated with each proliferating cell reaches its 
steady-state value of $(1-\rho)/\rho$. One may therefore infer that, at 
short times, the mean number of basal layer cells in clones arising from 
proliferating cells is given by 
\begin{eqnarray*}
\lim_{t\ll1/\Gamma} \langle n \rangle_{\rm pers.} = 1/\rho - (1/\rho-1)
e^{-\Gamma t}\,, 
\end{eqnarray*}
and that the early-time clone size distribution is Poisson-distributed, viz.
\begin{equation}
\label{eqn:shortTimeP}
\lim_{t\ll1/\Gamma} P^{\rm pers.}_{n\geq 2}(t) = \frac{\left(\langle n 
\rangle_{\rm pers.}-1\right)^{n-1}}{\left(e^{\langle n 
\rangle_{\rm pers.}-1}-1\right)(n-1)!}\,.
\end{equation}
%

%

\subsection{Fit to the data}
\begin{figure*}[ht]
\begin{center}
\includegraphics[width=7in]{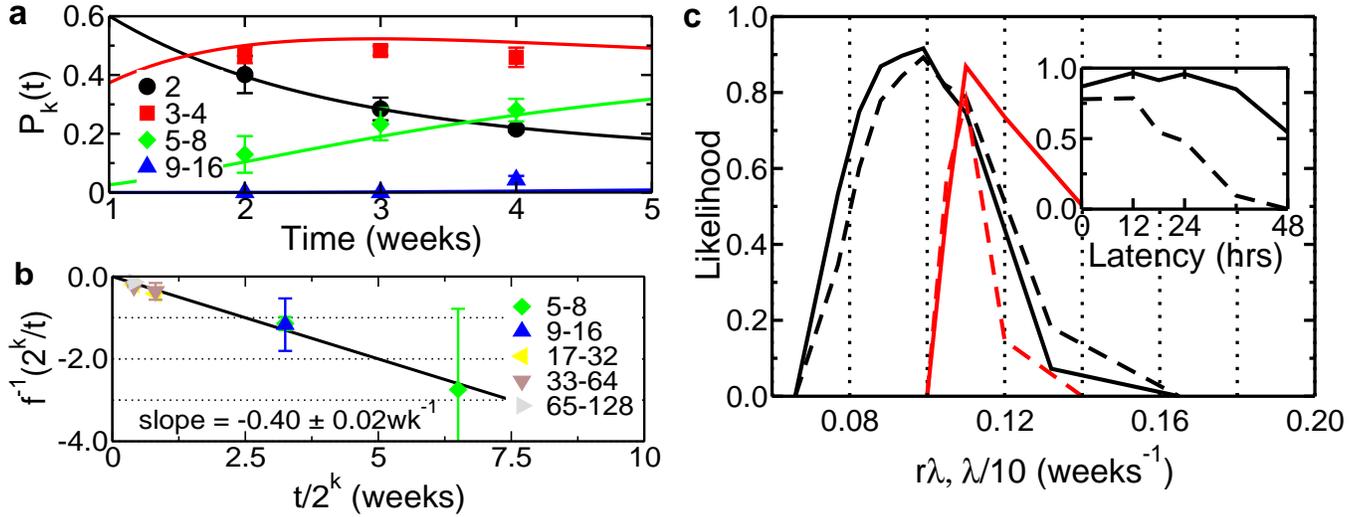}
\caption{\small 
(Color online)
(a) Fit of Eq.~(\ref{eqn:shortTimeP}) to the short-time clone size distributions.
At early times (a), the data is optimally fitted by Eq.~(\ref{eqn:shortTimeP}) using the value $\lambda=1.1$/week and the empirical value $\rho=0.22$ (solid lines show fit). To ensure integrity of the analysis, data for times earlier than week 2 have been excluded (see section~\ref{sec:EYFP}).
(b) Linearisation of the long-time asymptotic data using the ``inverse'' scaling function $f^{-1}(2^k/t)$ for $t\geq 13$ weeks and $k\geq3$ (see main text).
(c) Likelihood of the overall division rate $\lambda$ (red) and the symmetric division rate $r\lambda$ (black), as assessed from a $\chi^2$ test of the numerical solution to 
Eq.~(\ref{pMasterEqn})~\cite{numRecipes}. A fit to the basal-layer clone size distribution alone (dashed) is less discriminatory than a simultaneous fit to both the basal-layer and total clone size distributions (solid curves). The likelihood of $r\lambda$ is shown for the optimal value of $\lambda$, and vice-versa. \emph{Inset:} Referring to section~\ref{sec:stochastic}, the likelihood is plotted against the duration of a latency period ($\tau_{\rm min.}$) immediately following cell division, and assuming that division events are otherwise independent (see main text).
}
\label{fig:fit2Data}
\end{center}
\end{figure*}
With these insights it is now possible to attempt a fit of the model to the 
data. Referring to Fig.~\ref{fig:fit2Data}, one may infer the rate of cell division $\lambda$ from the short-time data, and the symmetric division rate $r\lambda$ from the long-time scaling data.  In particular, taking the fraction of proliferating 
cells in the basal layer to be $\rho=0.22$, a figure obtained experimentally 
by immunostaining using Ki67~\cite{Clayton:07}, a fit of Eq.~(\ref{eqn:shortTimeP}) 
to the short-time data (fig.~\ref{fig:fit2Data}(a)) is consistent with a transfer rate of 
$\Gamma=0.31$/week which, in turn, implies a rate of cell division of $\lambda=1.1$/week. 
Furthermore, by plotting the long-time, large-$k$,
size-distributions in terms of the ``inverse'' to the scaling function,
\begin{eqnarray*}
f^{-1}(2^k/t) & \equiv & \left( 2\ln\left[(1-(1-f(2^k/t))^{1/2})/2\right] \right)^{-1}\\
                  & = & \left( 2\ln\left[(1-(1-\mathcal{P}^{\rm pers.}_k(t))^{1/2})/2\right] \right)^{-1},
\end{eqnarray*}
the data converge 
onto a linear plot (Fig.~\ref{fig:fit2Data}(b)). The resulting slope takes the value $-r\lambda/\rho$, from which we may infer the symmetric division rate $r\lambda = 0.09\pm0.01$/week, and $r=0.08\pm 0.01$.

These figures compare well with an optimal 
fit of the \emph{entire} 
basal layer clone size distribution (Fig.~\ref{fig:scaling}), obtained by
numerically integrating the Master equation~(\ref{pMasterEqn}). 
The fitting procedure is shown in Fig.~\ref{fig:fit2Data}(c) (solid curves), where the likelihood of the model is evaluated for a range of values of $\lambda$ and $r\lambda$, as assessed from a $\chi^2$ test of the model solution~\cite{numRecipes}.
One may see that the likelihood is maximised with an overall division rate of $\lambda=1.1$/week and a symmetric division rate in the range $r\lambda=0.1\pm 0.01$/week, thus confirming the validity of the asymptotic fits. 
Moreover, the corresponding fit of both the basal layer distribution and the \emph{total} clone size distribution, including both basal and supra-basal cells, is equally favourable (Fig.~\ref{fig:fit2Data}(c), dashed). Thus, in the following sections we shall use the asymptotically fitted value of $r=0.08$, however any choice of the parameter in the range $r=0.08-0.10$ gives similar results.

Although the comparison of the experimental data with the model leaves little
doubt in its validity, it is important to question how discerning is the fit.
By itself, the observed increase in the size of persisting clones is sufficient
to rule out any model based on long-lived self-renewing stem cells, the basis
of the orthodox EPU model. However, could one construct a more complicated 
model, which would still yield a similar fit? Certainly, providing the 
long-time evolution is controlled by a single rate-determining process, the 
incorporation of further short-lived proliferating cell compartments (viz.
transit-amplifying cells) would not affect the observed long-time scaling 
behaviour. However, it seems unlikely that such generalisations would provide
an equally good fit to the short-time data. 

More importantly, it is crucial to emphasize that the current experimental 
arrangement would be insensitive to the presence of a small, quiescent, 
long-lived stem cell population. Yet, such a population could play a crucial 
role in \emph{non}-steady state dynamics such as that associated with wound healing or
development. We are therefore led to conclude that the range of clone fate
data for normal adult IFE are consistent with a simple (indeed, the simplest) 
non-equilibrium process involving just a single progenitor cell compartment.

\subsection{Stochastic behaviour of cell division}
\label{sec:stochastic}
\begin{figure*}[ht]
\begin{center}
\includegraphics[width=6.6in]{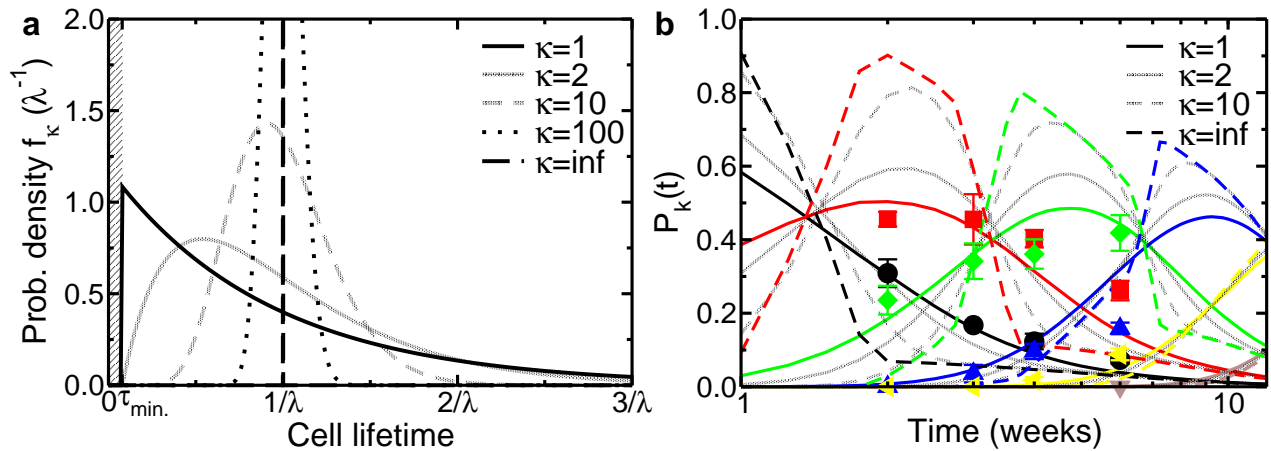}
\caption{\small 
(Color online).
(a) Examples of progenitor cell cycle-time distributions with the same average cycle time $1/\lambda$, ($\lambda=1.1$/week), and with a latency period of $\tau_{\rm min.}=12$ hours introduced between consecutive cell divisions (hashed region). The case $\kappa=1$ corresponds to a model of independent cell division such as assumed in section~\ref{sec:model}, but now accounting for an initial latency period.
 The case $\kappa\rightarrow \infty$ (black dashed) corresponds to all cells having an exact cell cycle-time of $1/\lambda$. Note that small values of $\kappa$ allow for both very short and very long cycle times.
(b) Using Monte-Carlo simulations of process (\ref{modelRateLaws}), the clone size distributions predicted by each of the different cycle-time distributions in (a) are compared with the empirical data. Data points show the size distribution of persisting clones including supra-basal layer cells over the first 6 weeks post-labelling (extracted from data given fully in Ref.~\cite{Clayton:07}; for legend see Fig.~\ref{fig:scaling}), and the theoretical curves correspond to the same legend as in (a). All of the models give an optimal fit with the same value of $\lambda=1.1$/week, $r=0.08$.
}
\label{fig:stochastic}
\end{center}
\end{figure*}
At this stage, it is useful to reflect upon the sensitivity of the model to
the stochasticity assumption applied to the process of cell division. 
Clearly,
the scaling behaviour (Eq.~\ref{P_Exact_All_Cells}) depends critically on the 
statistical independence of successive cell divisions; each cell division 
results in symmetric/asymmetric cell fate with relative probabilities as 
detailed in~(\ref{modelRateLaws2}).
But, to what extent would the findings above 
be compromised if the cell \emph{cycle-time}, i.e. the time between consecutive cell divisions, were not determined by an independent stochastic process?
This question may have important ramifications, because the assumption of
independent cell division, used in formulating the Master equation~(\ref{pMasterEqn}), introduces a manifestly unphysical behaviour by allowing cells to have arbitrarily short cycle times. 
Moreover, although a wide distribution of cell cycle-times has been observed for human keratinocytes \emph{in vitro}~\cite{DoverPotten:88}, it is possible that that keratinocytes \emph{in vivo} may divide in \emph{synchrony}, giving
a cell cycle-time distribution narrowly centered about the mean ($1/\lambda$).
In the following, we shall address both of these points: 
Firstly, we shall show that, up to some potential latency period (the time 
delay before a newly-divided cell is able to divide again), consecutive
cell divisions occur independently as an asychronous, Poisson process. 
Secondly, while the data is insufficient to detect a latency period of $12$ 
hours or less between consecutive cell divisions, the data does discriminates 
against a period lasting longer than $24$ hours. 

To investigate the degree to which the model is sensitive to 
the particular 
cell cycle-time distribution, 
let us revisit the original model of independent 
cell division with several variations:
Firstly, we introduce a latency period of $\tau_{\rm min}$ 
immediately following cell division, 
in which daughter cells cannot divide. This biologically-motivated constraint 
renders a more complicated yet more realistic model of cell division than the
 idealised system studied in the previous section. 
Motivated by observations of the minimal cycle-time of (human) keratinocytes~\cite{DoverPotten:88}, where a latency period of $\tau_{\rm min}\simeq10$ hours was observed \emph{in vitro}, we shall here consider the a range latency periods of up to $48$ hours.
%
Secondly, we compare the empirical clone size distributions 
with a model where
all progenitor cells have a cycle-time of exactly $1/\lambda$, i.e.
where cells within each clone divide in perfect synchrony.
Finally, we shall investigate a range of intermediate models 
with different distributions 
of progenitor cell cycle-time (see Fig.~\ref{fig:stochastic}(a)).

Technically, the resulting clone size distributions may be evaluated 
through Monte Carlo simulations of the non-equilibrium process (\ref{modelRateLaws})
with the cycle-time $\tau$ of each proliferating cell selected at
random from a Gamma distribution of the form
$$f_{\kappa}(\tau) = \left\{\begin{array}{cc}
0 & \tau<\tau_{\rm min.} \\
 \frac{\kappa^\kappa (\tau-\tau_{\rm min.})^{\kappa-1}}{\bar{\tau}^\kappa\Gamma(\kappa)}e^{-\frac{\kappa(\tau-\tau_{\rm min.})}{\bar{\tau}}} & \tau\ge \tau_{\rm min.}\end{array}\right.,$$
where $\bar{\tau} = 1/\lambda - \tau_{\rm min.}$ is the average time to division following the initial latency period $\tau_{\rm min.}$, and $\kappa$ is the ``shape parameter'' of the Gamma distribution. In particular, the choice of shape parameter $\kappa=1$ corresponds to the exponential distribution which characterises the independent cell cycle-time distribution, whereas $\kappa \rightarrow\infty$ describes the case in which all A-cells have an exact cycle-time of $1/\lambda$ (see Fig.~\ref{fig:stochastic}(a)). Then, to reflect the assumption that initially-labelled, spatially separated, progenitor cells have uncorrelated cell cycles, the time to the initial division event post-labelling is adjusted by a random time $\tau\in[0,1/\lambda]$. 
Finally, for an unbiased comparison of the models, we optimise the value of $\lambda$ for each model separately against the empirical data, whilst keeping $r\lambda = \rm const.$ to ensure an optimal fit of the long-time data, as discussed below.

The resulting clone size distributions are shown in Fig.~\ref{fig:stochastic}(b), where the case of independent division following a 12-hour latency ($\kappa=1$) and the exact cycle-time case ($\kappa\rightarrow\infty$) are compared to the empirical \emph{total} clone size distribution, which includes both basal and supra-basal (type C) cells, over the first 6 weeks post-labelling.
Two intermediate cases are also shown for comparison ($\kappa=2, 10$). 
Focusing first on the results for the case $\kappa=1$, which bears closest resemblance to the Markovian model analysed using the Master Equation (\ref{pMasterEqn}), one may see by inspection that the quality of the fit to the data remains good even when the effects of a latency period between cell divisions is taken into account. More rigorously, a likelihood analysis reveals that the two cases are statistically indistinguishable (see Fig.~\ref{fig:fit2Data}(c), inset), which indicates that the duration of a latency period of $\tau_{\rm min.} \lesssim 12$ hours is beyond the current empirical resolution. However, referring to Fig.~\ref{fig:fit2Data}c (inset), a similar analysis of longer latency periods reveals that for periods of $\tau_{\rm min.}\gtrsim 24$ hours, the fit to the data is significantly poorer.

Turning next to the predicted basal-layer clone size distributions at late times ($t\gtrsim\rho/r\lambda$) (not shown), one may see that all of the proposed distributions asymptotically converge: Starting with exactly one cell, then the moment-generating function $G(q,s) = \sum_{n=0}^{\infty}p_n(s)q^n$ associated with the A cell population distribution $p_n(s)$ after $s$ cell cycles satisfies the recursion relation~\cite{KarlinTaylor:75}:
$$G(q,s+1) - G(q,s) = r\left( G(q,s)-1\right)^2\, ,$$
which asymptotes to the continuous master equation
$\lim_{s\gg1} \partial_s G(q,s) = r\left( G(q,s)-1\right)^2\, ,$
with the relative magnitude of the leading-order correction dropping off as $1/s$. 
But with $s=\lambda t$, this equation is simply the master equation for the moment-generating function associated with the original model, Eq.~\ref{P_Exact_ACells}, and so the two models converge. One may therefore conclude that, beyond the first several weeks of the experiment ($t\gg1/\lambda$), the fit to the data is sensitive only to the average cycle time of progenitor cells. With this in mind, we note that for the case of perfectly synchronous cell division, 
an optimal (albeit poor) numerical fit was obtained when $\lambda=1.2$/week, a figure that compares well with the fit for the independent case. It appears therefore that \emph{the predicted average cell division rate ($\lambda$) is insensitive to the shape of the cell cycle distribution}.

Finally, let us turn to the early time behaviour ($t\sim1/\lambda$), where the predicted
distributions are distinct. Referring to Fig.~\ref{fig:stochastic}(b), one may see, at 2-4 weeks post-labelling, that relatively large clones ($5-8$ cells) appear earlier than expected by a model assuming synchronous division,
and that, compared with the same model, a sizeable proportion of small clones (e.g., $2$ cells) lingers on for far longer than expected.
The same behaviour is observed for the basal layer clone size distribution (not shown).
One may therefore infer that cell division conforms to a model of \emph{independent} rather than \emph{synchronous} division, allowing for some progenitor cells to divide unusually early, and for others to remain quiescent for an unusually long period of time.  

In summary, we have established that, following division, progenitor cells do not divide for a period that is likely to last up to $12$ hours, and not more than $24$ hours. After this latency period, the data is consistent with cells switching to a mode of independent, asynchronous, cell division. These results shed light on why the simple model of independent cell division presented in section~\ref{sec:model} succeeds in producing such a remarkable fit to the data.

\subsection{Labelling efficiency and EYFP accumulation in basal cells}
\label{sec:EYFP}

Although the integrity of the fit of the model to the data provides some 
confidence in its applicability to the experimental system, its viability as 
a model of epidermal homeostasis rests on the labelled clone population being 
representative of all cells in the IFE. Already, we have seen that the 
model, and by inference, the labelled clone population, has the capacity to
self-renew. However, the slow accumulation of EYFP after induction, together
with the question of the relative labelling efficiency of the two basal layer cell types, leaves 
open the question of the very short-time behaviour. Accepting the validity of 
the model, we are now in a position to address this regime. 

In doing so, it is particularly useful to refer to the time evolution of 
clone size as measured by the average number of basal cells in a persisting 
clone. As expected from the scaling analysis discussed in 
section~\ref{sec:scaling}, a comparison of the experimental data with that 
predicted by the proposed cell kinetic model shows a good agreement at long 
times (Fig.~\ref{fig:avgSize}). However, comparison of the data at 
intermediate time-scales provides significant new insight. In particular, if 
we assume \emph{equal} labelling efficiency of progenitor and differentiated 
cells, i.e. that both cell types label in proportion to their steady-state 
population (shown as the lower (red) curve in the Fig.~\ref{fig:avgSize}), 
then there is a 
substantial departure of the predicted curve from the experimental data for 
times of between two and six weeks. Intriguingly, if we assume that 
differentiated cells simply don't label, then the agreement of the data with 
theory is excellent from two weeks on! We are therefore lead to conclude that, 
at least from two weeks, all labelled clones derive from progenitor cells 
labelled at induction.

With this in mind, we may now turn to the average clone size as inferred 
from the data at two days and one week. Here one finds that the model appears 
to substantially over-estimate the clone size. Indeed, Fig.~\ref{fig:avgSize} 
suggests that the average clone size is pinned near unity until beyond the 
first week post-labelling, i.e. the relative population of single-cell clones 
is significantly \emph{larger} than expected at one week, yet falls 
dramatically to the theoretical value at two weeks. Referring again to the 
slow accumulation of EYFP, can one explain the over-representation of 
single-cell clones at one week post-labelling? At one week, two-cell clones 
are observed soon after cell division, and thus express lower concentrations 
of EYFP compared to single-cell clones. As a result they may be 
under-represented. At later times, all labelled clones become visible as EYFP 
concentration grows, explaining the coincidence of experiment and theory at 
two weeks. It follows, of course, that the size distributions at later time 
points would be unaffected by slow EYFP accumulation. However, a full 
explanation of this effect warrants further experimental investigation, and 
is beyond the scope of this paper. 

\section{Manifestation of mutations in clonal distributions}

Having elucidated the mechanism of normal skin maintenance, it is interesting 
to address its potential as a predictive tool in clonal analysis. 
Conceptually, the action of mutations, drug treatments or other environmental 
changes to the tissue can effect the non-equilibrium dynamics in a variety of 
ways: Firstly, a revision of cell division rates or ``branching ratios'' (i.e. 
symmetric vs. asymmetric) of \emph{all} cells may drive the system towards 
either a new non-equilibrium steady-state or towards a non-steady state 
evolution resulting in atrofication or unconstrained growth of the tissue. 
(The development of closed non steady-state behaviour in the form of limit 
cycles seems infeasible in the context of cellular structures.) Secondly, 
the stochastic revision of cell division rates or branching ratios of 
\emph{individual} cells may lead to cancerous growth or extinction of a 
sub-population of clones. The former may be referred to as a ``global 
perturbation'' of the cell division process while the second can be referred 
to as ``local''. In both cases, one may expect clonal analysis to provide a 
precise diagnostic tool in accessing cell kinetics. To target our discussion 
to the current experimental system, in the following we will focus on the 
action of a local perturbation in the form of a carcinogenic mutation, 
reserving discussion of a global perturbation, and its ramifications for the 
study of drug treatment, to a separate publication.

Let us then consider the action of a local perturbation involving the 
activation of a cancer gene in a small number of epidermal cells, which leads 
to the eventual formation of
tumours. In the experimental system, 
one can envisage the treatment coinciding with label induction, for example by 
simultaneously activating the EYFP and the cancer gene. In this case, clonal 
fate data should simply reflect a modified model of cell proliferation leading 
to the eventual failure of the steady-state model of tissue maintenance. 

\subsection{A simple model of carcinogenesis}

To quantify the process of cancer onset, we start by establishing the simplest 
possible changes to process~(\ref{modelRateLaws}) which may be associated with 
tumour growth. 
Cancer is widely held to be a disease caused by genetic instability that is thought 
to arise when a progenitor cell undergoes a series of mutations~
\cite{Fearon:87, Fearon_Vogelstein:90, Keller:78}. 
As a result, cells within the mutant clone prefer to proliferate, on average, 
over processes 
leading to terminal differentiation or death. In this investigation we shall 
consider a ``simple'' cancer resulting from \emph{two} rate-limiting 
mutations: Referring to our proposed labelling experiment, the controlled 
induction of a cancer-causing mutation during label induction defines the
first mutation; a second, rate-limiting step then occurs with the stochastic 
occurrence of a second cancer causing mutation.  
Examples of the first type of mutation may be genes that affect the ability 
of a cell to respond to genetic changes of the cell, e.g. {\it p53}, whilst the 
second mutation may be of a gene that affects clone fate such as the 
{\it Ras} oncogene \cite{Fearon_Vogelstein:90}. 
We may therefore distinguish between ``stage one'' mutated 
cells, which maintain the steady-state, and ``stage two'' cells, which have 
the capacity for tumour formation. 

The resulting process of cell proliferation is set by three parameters: The 
overall rate of mutation $\nu$ from a stage one A cell into a cancerous stage 
two cell; the division rate $\mu$ of the stage two cells; and the degree of 
imbalance $\Delta$ between their stochastic rate of proliferation and 
differentiation. In summary, focusing on the proliferating cell compartment 
only, and denoting the stage two mutated cells as type A$^*$, then the 
revised cell proliferation model includes 
the additional non-equilibrium processes
\begin{equation}
\label{processEqn:Cancer}
\begin{array}{lcl}
A &  \stackrel{\nu} { \longrightarrow} & A^*\\
A^* &  \stackrel{\mu} { \longrightarrow}  &
        \left\{\begin{array}{cl}
        A^*+A^* & {\rm\ prob.\ } (1+\Delta)/2 \\
        \emptyset &  {\rm\ prob.\ } (1-\Delta)/2\,. \\
\end{array}\right. 
\end{array}
\end{equation}
The rate $\nu$ may be interpreted as the mean rate with which a stage-one cell 
acquires an additional mutation necessary to activate a second oncogene. The 
mutated cells then give rise, on average, to an exponentially growing cell 
lineage with growth rate $\Delta\mu$.
 
This nonequilibrium process was originally addressed by Kendall, who predicted 
the distribution in the number of tumours detected at time $t$ after 
mutation~\cite{Kendall:60}. His focus on tumour statistics may reflect the 
experimental limitations in clonal analysis at the time: Until recently it 
was not possible to reliably detect clones at all, let alone to count the 
number of cells per clone. Experimentally, however, the clone size 
distributions are a more efficient measure of cell kinetics than the tumour 
number distributions, because they result in a far richer data set, and are 
accessible within weeks rather than months. We shall therefore extend 
Kendall's approach to predict the clone size distributions at times far 
earlier than tumour appearance.
\begin{figure*}[t]
\centering
\includegraphics[width=7in]{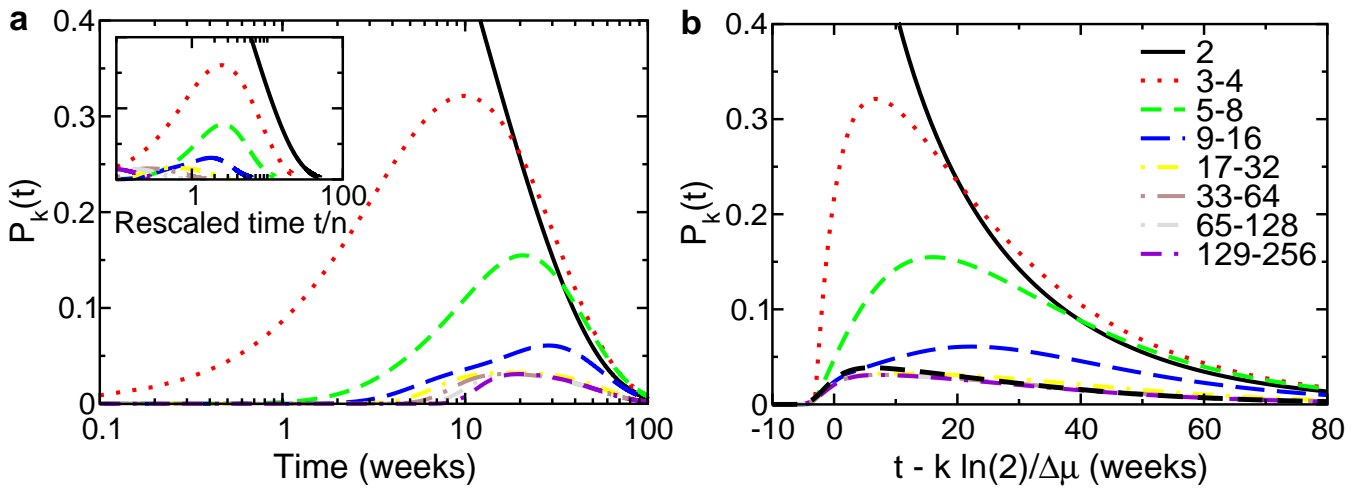}
\caption{\small 
(Color online)
(a) The total number of basal layer cells per labelled clone 
during the onset of cancer according to process~(\ref{processEqn:Cancer}). The 
figure was plotted by numerically integrating Eq.~(\ref{eqn:cancerMaster}) 
using the empirical value $r\lambda = 0.088{\rm/week}$ found for normal skin, 
together with hypothetical values of the cancer growth parameters $\nu=0.1\, 
r\lambda,\ \mu=10\, r\lambda$, and $\Delta= 0.5$. To compare with normal skin, 
the predicted clone size distributions are replotted against the rescaled time 
coordinate $t/2^k\mapsto t$ in (a) inset. In contrast with 
Fig.~\ref{fig:scaling}, here the curves fail to converge. In (b), the same 
curves are shown converge onto the universal form given in 
Eq.~(\ref{cancerSummarySoln}) (dashed) when they are plotted against a new 
rescaled time $t\mapsto t'_k = t + k\ln2/\Delta\mu$. Note that the large-clone 
distributions converge rapidly, whereas the distributions for smaller clones are affected by the 
non-negligible contribution of non-cancerous (A) cells to the small-clone 
size distribution.}
\label{fig:cancer}
\end{figure*}

\subsection{Clonal behaviour during early-stage cancer}
\label{sec:cancer:Soln}
To familiarise ourselves with the modified model, consider the evolution of 
the average clone size with time. Focusing on the proliferating cell 
compartment with $n$ type A cells and $n^*$ type A$^*$ cells in a clone, the relevant mean-field equations are
\begin{eqnarray*}
&&\partial_t \langle n \rangle = -\nu \langle n \rangle\,, \\
&&\partial_t \langle n^* \rangle = \nu\langle n \rangle + \Delta\mu 
\langle n^* \rangle\,,
\end{eqnarray*}
which give the expected shift from linear growth of clones in normal skin to 
that of exponential growth, $\langle n+n^*\rangle = (\nu e^{\Delta\mu t} + \Delta\mu e^{-\nu t})/(\nu + \Delta\mu)$.
More interestingly, referring to the Master equation below, one may show that 
the variance in clone size also changes qualitatively: Whereas for normal 
skin the RMS variance in clone size grows as $t^{1/2}$, here the variance in 
the long-time limit is \emph{finite},
\begin{eqnarray*}
\lim_{t\rightarrow \infty} \frac{\langle (n^*-\langle n^* \rangle)^2
\rangle^{1/2}}{\langle n^* \rangle} = \sqrt{1 + \Delta^{-1} }\,.
\end{eqnarray*}
That is, the relative broadening of the clone size distribution observed in 
normal skin is halted by the introduction of an exponentially growing cell 
population.

These observations may already provide a crude method for identifying 
carcinogenesis through clonal analysis. To do better, it becomes necessary to 
solve for the full size distribution by extending the Master 
equation~(\ref{pMasterEqn}) to include process (\ref{processEqn:Cancer}). If 
we neglect the fate of differentiated cells, then the Master equation now 
describes the evolution of the probability $P_{n,n^*}(t)$ for finding $n$ 
type A cells and $n^*$ type A$^*$ cells in a clone,
\begin{widetext}
\begin{eqnarray}
\label{eqn:cancerMaster}
&&\partial_t P_{n,n^*} = r\lambda\left[(n-1)P_{n-1,n^*} - n P_{n,n^*} 
\right] + r\lambda\left[(n+1)P_{n+1,n^*}-n P_{n,n^*} \right]+ 
\nu\left[(n+1)P_{n+1,n^*-1}-n P_{n,n^*} \right]\nonumber\\
&& \qquad\qquad + \frac{1+\Delta}{2}\mu\left[(n^*-1)P_{n,n^*-1} - 
n^* P_{n,n^*} \right]+ \frac{1-\Delta}{2}\mu\left[(n^*+1)P_{n,n^*+1} 
- n^* P_{n,n^*} \right]\,,
\nonumber
\end{eqnarray}
\end{widetext}
subject to the experimental boundary condition $P_{n,n^*}(0) = \delta_{1,0}$ 
corresponding to exactly one ``stage one'' cell per clone at $t=0$. As for 
the case of normal skin, we shall later be interested in the distribution of 
persistent clones, defined as, 
\begin{eqnarray*}
\mathcal{P}^{\rm(canc.)}_{2^k}(t) = \sum_{N=2^{k-1}+1}^{2^k} \sum_{n=0}^{N} 
\frac{P_{n,N-n}(t)}{1-P_{0,0}(t)-P_{1,0}(t)-P_{0,1}(t)}\,.
\end{eqnarray*}
While it is not possible to solve Eq.~(\ref{eqn:cancerMaster}) analytically, 
progress may be made when we allow for the widely-accepted view that 
tumours are \emph{monoclonal}, that is 
they arise from a single ``stage two'' mutated cell~\cite{Fearon_Vogelstein:90}. 
This assumption conveniently limits us to 
the parameter space $\nu\ll \Delta\mu$, for which an approximate long-time 
solution for the full clone size distribution may be found. 

Referring to the appendix for details, we find that the binned clone size 
distribution takes the long-time asymptotic scaling form, 
\begin{eqnarray}
\label{cancerSummarySoln}
\mathcal{P}^{\rm(canc.)}_{k}(t) \simeq
\mathcal{N}\left[ \mathcal{I}_{\beta,a}\left(\frac{1}{2\phi_k(t)}\right) - 
\mathcal{I}_{\beta,a}\left(\frac{1}{\phi_k(t)}\right) \right]\,,
\end{eqnarray}
where $\phi_k(t)=(1+\Delta^{-1}) e^{\Delta\mu t}/2^k$
\begin{eqnarray*}
\mathcal{I}_{\beta,a}(x)=\int_1^\infty d\zeta \frac{\zeta^{-1-\beta} 
e^{-x\zeta} }{(1 + a\zeta^{-\beta})^2}\,,
\end{eqnarray*}
$\mathcal{N}=\frac{4r\lambda \chi^2} { \Delta \mu (\chi + \nu/2r\lambda)}$,
$\chi^2=\left(\frac{\nu}{2r\lambda}\right)^2 +2\frac{\Delta\nu}{(1+\Delta)
r\lambda}$, $\beta=2\chi r\lambda/\Delta\mu$, and $a=\frac{2r\lambda\chi-
\nu}{2r\lambda\chi+\nu}$. Despite its apparent complexity, this distribution 
is characterised by a simple scaling behaviour: Referring to 
Fig.~\ref{fig:cancer}(a), the predicted clone size distributions are plotted 
using the scaling appropriate to the normal (unperturbed) system (cf. 
Fig.~\ref{fig:scaling}). In this case, it is apparent that the scaling 
$t\mapsto t/2^k$ fails. By contrast, from the expression for $\phi_k(t)$, it
is clear that the size distributions should scale according to the time 
translation, $t\mapsto t'_k = t + k\ln2/\Delta\mu$ as confirmed by the 
results shown in Fig.~\ref{fig:cancer}(b). 
Further consideration of the size distribution exposes several additional 
features, which may provide further access to the new model parameters:\\

\begin{itemize} 

\item The long-time distribution decays with a rate $\beta \Delta \mu$:

\end{itemize}

\noindent
Expanding $\mathcal{I}_{\beta,a}(x)$ for small $x$ gives us the asymptotic 
form of the universal decay curve. For $\beta<1$, consistent with the 
monoclonicity requirement $\Delta\mu\gg\nu$, we find
\begin{equation}
\label{eqn:asymptotic}
\lim_{t\gg\Delta\mu} \mathcal{P}^{\rm(canc.)}_{k}(t) = \mathcal{N}\,
\Gamma(-\beta)(2^{-\beta}-1)\phi_k(t)^{-\beta},
\end{equation}
where $\Gamma(x)$ denotes the Gamma function. This expression allows us to 
estimate $\beta$ from the rescaled clone size distributions, providing access 
to the cell division and mutation parameters of the observed cells.

\begin{itemize}

\item The probability of tumour formation is finite: 

\end{itemize}

\noindent This is a well-known feature of the simple non-critical birth-death 
process~(\ref{processEqn:Cancer})~\cite{Bailey}. Referring to the appendix, we 
find that the probability $p_T$ for any given clone to survive and form a 
tumour is finite,
\begin{eqnarray*}
p_T 
& = & 1 + \frac{\nu}{2r\lambda} - \sqrt{\left( \frac{\nu}{2r\lambda}\right)^2 
+ \frac{2\Delta\nu}{r\lambda(1+\Delta)}}\,.
\end{eqnarray*}

As a result, the onset of cancer will halt the steady decrease in the density 
of labelled clones that is a hallmark of the unperturbed system.

These properties, and especially the change in scaling behaviour, allow the 
onset of early-stage cancer to be identified from observations of clones less 
than one hundred cells in size. This may provide a dramatic improvement both 
in speed and accuracy over current experimental models, which rely on much 
later observations of tumours (or hyperplasias) in order to deduce the cell 
kinetics at early-stages.

\section{Conclusions}

To summarize, we have shown that the range of clone fate data obtained from 
measurements of murine tail epidermis are consistent with a remarkably simple 
stochastic model of cell division and differentiation involving just one 
proliferating cell compartment. These findings overturn a long-standing 
paradigm of epidermal fate which places emphasis on a stem cell supported 
epidermal proliferative unit. As well as providing significant new insight
into the mechanism of epidermal homeostasis, these results suggest the 
utility of inducible genetic labelleling as a means to resolve the mechanism
of cell fate in other tissue types, and as a means to explore quantitatively
the effects of drug treatment and mutation. 

To conclude, we note that the analysis above has focused on the dynamics of 
the clonal population without regard to the spatial characteristics. Indeed, 
we have implicitly assumed that any model capable of describing the cell size 
distributions will also succeed in maintaining the near-uniform areal cell 
density observed in the basal layer. However, it is known that, when augmented 
by spatial diffusion, a simple Galton-Watson birth-death process leads to 
``cluster'' formation in the two-dimensional system whereupon local cell 
densities diverge logarithmically~\cite{Houchmandzadeh:02,Young:01}. 
Significantly, these divergences can not be regulated through a 
density-dependent mobility. Understanding how the Galton-Watson process 
emerges from a two-dimensional reaction-diffusion type process represents a 
significant future challenge. 

From a practical perspective, there is also the significant question of how 
the cell kinetic model might be generalised to describe other forms of epidermis.
In particular, it is not feasible to repeat these experiments \emph{in vivo} 
in human epidermis, a system of obvious medical significance. Therefore, 
it may be of great interest to determine, in future studies, the extent to 
which our results compare with the behaviour found in other systems. 

Lastly, our analysis of the cancer system referred to the relatively simple 
case of a two-stage mutation. It is, of course, well-known that tumour 
formation is usually the result of multiple mutations. Understanding whether
clonal fate data can be used to probe the kinetics of \emph{multi-stage} 
mutation remains an interesting future challenge. 

{\sc Acknowledgements:} We are grateful to Sam Edwards, Martin Evans, Bill 
Harris, Marc Kirschner, and Gunter Sch\"utz for useful discussions. 
The experimental aspect of this work was funded by the Medical Research Council, 
Association for International Cancer Research and Cancer Research UK.


\appendix 

\section*{Appendix: Clone size distributions in the two-stage cancer model}

To derive the clone size distribution given in Eq.~(\ref{cancerSummarySoln}), 
we start by quoting the known result for the probability distribution 
$\Pi_{n^*}(t;\tau)$ of finding $n^*$ stage-two A$^*$ cells at time $t$ 
starting from a single A$^*$ cell at time $\tau$~\cite{Bailey},
\begin{widetext}
\begin{equation}
\label{cancer_P_soln}
\Pi_{n^*}(t;\tau) =\left\{\begin{array}{ll}
\frac{(1-\Delta)\left(1-e^{-\Delta\mu (t-\tau)}\right)}{(1+\Delta)-
(1-\Delta)e^{-\Delta\mu (t-\tau)}} & {\rm for\ } n^*=0 \\
\left(\frac{2\Delta}{(1+\Delta)\left(1-e^{-\Delta\mu (t-\tau)}\right)}
\right)^2 e^{-\Delta\mu (t-\tau)} \left( 1 - \frac{2\Delta e^{-\Delta\mu 
(t-\tau)}}{(1+\Delta)-(1-\Delta)e^{-\Delta\mu (t-\tau)}}\right)^{n^*-1} \qquad
& {\rm for\ } n^*\geq 1
\end{array}\right.
\end{equation}
\end{widetext}
When $\Delta\mu(t-\tau)\gg1$, this distribution asymptotes to the form
\begin{equation}
\label{cancer_P_Approx_soln}
\Pi_{n^*}(t;\tau) \simeq
\left\{\begin{array}{ll}
\frac{(1-\Delta)}{(1+\Delta)} & {\rm for\ } n^*=0 \\
\\
\left(\frac{2\Delta}{(1+\Delta)}\right)^2 e^{-\Delta\mu (t-\tau)}\times \\
\ \ \ \  \exp\left(-n^*\frac{2\Delta}{(1+\Delta)}e^{-\Delta\mu (t-\tau)}
\right) & {\rm for\ } n^*\geq 1
\end{array}\right.
\end{equation}
From the value of $\Pi_{0}$ we see that even when a cell has mutated, it is 
not guaranteed to result in a tumour: This will only occur with a probability 
of $f = 1-\Pi_0(t\rightarrow\infty) = (2\Delta)/(1+\Delta)$. The value of $f$ 
plays an important role in determining the statistics of tumour formation, as 
will be seen below.

We now make two approximations: First, we take the long-time clone size 
distribution to be dominated by the statistics of A$^*$ cells. This is a safe 
assumption at times $t\gtrsim 1/\nu$ and $t\gg 1/\Delta\mu$, as may be seen by 
considering the behaviour of the mean-field equations in 
section~\ref{sec:cancer:Soln}. This approximation allows us to focus on the 
size distribution of A$^*$ cells only, $p_{n^*}(t)$, which is related to the 
full clone size distribution by the sum $p_{n^*}(t)=\sum_{n=0}^{\infty} 
P_{n,n^*}(t)$. Secondly, we assume that the entire population of type A$^*$ 
cells in each clone arises from the first mutated cell that gives rise to a 
stable, exponentially growing lineage of cells. This corresponds to the 
condition $\nu\ll \Delta\mu$, as discussed in the main text. 

With these two approximations, the probability of finding a labelled clone 
containing $n^*>0$ mutated cells is given by the population distribution of 
the \emph{first surviving cell lineage} of A$^*$ cells,
\begin{eqnarray}
\label{eqn:Case2PDefn}
p_{n^*}(t) \simeq \mathcal{N} \int_0^t  \Pi_{n^*}(t-\tau) \sum_{m=1}^{\infty} 
(1-f)^{m-1} r_m(\tau) d\tau\,,
\end{eqnarray}
where $\mathcal{N}$ is some normalisation constant, and we have introduced 
the probability $r_m(\tau)d\tau$ for the $m$-th lineage of mutated cells 
within a given clone to be created during the interval $\tau\rightarrow \tau
+d\tau$ through the mutation process $A\rightarrow A^*$. The weight factor 
$(1-f)^{m-1}$ gives the probability that the first $m-1$ cell lineages of 
A$^*$ cells within a clone will become extinct --- a situation necessary to 
make the $m$-th cell line relevant to the distribution according to the 
monoclonal approximation.

The rates $r_m(\tau)$ may be accessed by considering the probability 
$w_{n,m}(t)$ that a clone containing $n$ type A cells at time $t$ also 
contains $m$ \emph{independent lineages} of mutated A$^*$ cells, each arising 
from a separate mutation event. (Later we shall treat the evolution of each of 
these cell lines post-creation). To solve for $w_{n,m}(t)$ we must introduce 
its moment-generating function $G(q,Q^*; t) \equiv \sum_{n=0}^\infty
\sum_{m=0}^\infty w_{n,m}(t) q^n (Q^*)^{m}$, which evolves (from 
Eq.~\ref{eqn:cancerMaster}) according to the dynamical equation,
\begin{equation}
\label{eqn:GenFnEqn1}
\dot{G} = \left[r\lambda(q-1)^2+\nu(Q^*-q)\right]\partial_qG\,.
\end{equation}
Solving this equation subject to the initial condition of one ``stage-one'' 
(A) cell per clone, we find the solution
\begin{eqnarray}
\label{G1Soln}
&&G(q,Q^*; t)=\xi_{Q^*}  - \frac{2 \xi_{Q^*} }{1+\frac{ \xi_{Q^*}+(q-1
-\frac{\nu}{2r\lambda})}{ \xi_{Q^*}-(q-1-\frac{\nu}{2r\lambda})}e^{-2 \xi_{Q^*} 
r\lambda t}} \nonumber \\
&&\qquad\qquad\qquad\qquad\qquad + 1 + \frac{\nu}{2r\lambda}\,,
\end{eqnarray}
with $\xi_{Q^*} \equiv \sqrt{(\frac{\nu}{2r\lambda})^2 + \frac{\nu}{r\lambda}(1-
Q^*)}$. Eq.~(\ref{G1Soln}) describes the evolution of a single A cell as it 
proliferates and eventually gives rise to a set of internal lines of mutated 
cells. 

Before we proceed to find $p_{n^*}(t)$, note that setting $q=1,Q^*=1-f$ in 
Eq.~(\ref{G1Soln}) gives us the result (quoted in the main text) for the 
asymptotic fraction $p_T$ of clones in which all mutated cell lines become 
extinct, 
\begin{eqnarray*}
p_T = 1 + \frac{\nu}{2r\lambda} - \sqrt{\left(\frac{\nu}{2r\lambda}\right)^2+
\frac{\nu f}{r\lambda}}\,.
\end{eqnarray*}

On the other hand, setting $q=1$ only in Eq.~(\ref{G1Soln}) gives the 
moment-generating function for (yet another) distribution $W_{m}(t)  = 
\sum_{n=0}^\infty w_{n,m}(t)$ of a clone containing $m$ independent lines of 
A$^*$ cells irrespective of the number of normal cells in the clone. Finally, 
noting that $\dot{W}_{m}(t) = r_m - r_{m+1}$ then gives:
\begin{equation*}
\label{A*RateEqn}
r_m(t) = - \sum_{n=0}^{m-1} \dot{W}_{m}(t)\,,
\end{equation*}
which we may substitute into Eq.~(\ref{eqn:Case2PDefn}) to find (for $n^*>0$)
\begin{eqnarray}
\label{eqn:Case2PSimplified1}
p_{n^*}(t) \simeq - \mathcal{N} \int_0^t d\tau  \Pi_{n^*}(t-\tau) 
\dot{G}(1, 1-f; \tau)\,.
\end{eqnarray}
From this expression, simplified by the large-$n^*$ approximation 
$(\sum_{n^*} \simeq \int dn )$, we obtain the final form of the binned size 
distribution given in Eq.~(\ref{cancerSummarySoln}).


\end{document}